\documentclass{ws-procs9x6}

\def\Journal#1#2#3#4{{#1} {\bf #2}, #3 (#4)}
\def\EPJA{{\em Eur. Phys. J.} A}

\def\NIMA{{\em Nucl. Instr. Meth.} A}
\def\NPA{{\em Nucl. Phys.} A}
\def\NPB{{\em Nucl. Phys.} B}
\def\PLA{{\em Phys. Lett.} A}
\def\PLB{{\em Phys. Lett.} B}

\def\PPNP{\em Prog. Part. Nucl. Phys.}
\def\PRL{\em Phys. Rev. Lett.}
\def\PREV{\em Phys. Rev.}

\def\PRD{{\em Phys. Rev.} D}
\def\PRC{{\em Phys. Rev.} C}

\def\ZPA{{\em Z. Phys.} A}

\begin{document}

\newcommand{\Den}{$D(\vec e,e'\vec n)$~}
\newcommand{\Hen}{$^{3}$$\vec{\mbox{He}}(\vec e,e'n)$~}

\title 
      {Electromagnetic form factors of the nucleon}


\author{Hartmut Schmieden}

\address{Physikalisches Institut \\
         Nussallee 12            \\
         D-53115 Bonn, Germany   \\
         E-mail:  schmieden@physik.uni-bonn.de}

\maketitle

\abstracts{
Elastic form factors provide information about the low energy structure
of composite particles.
Recent double polarization coincidence experiments significantly 
improved our knowledge of
proton and neutron form factors.
Recoil polarization measurements in the $p(\vec e, e'\vec p)$
reaction proved that 
at momentum transfers above $Q^2 \simeq 1.5$\,(GeV/c)$^2$  
the electric form factor of the proton 
falls significantly faster than the dipole expectation.
The close--to--dipole shape at low $Q^2$ of the neutron magnetic form factor
is now confirmed by independent measurements.
For the neutron electric form factor \Hen, $\vec D(\vec e,e'n)$  
and \Den double polarization experiments have provided model independent 
results, within their statistical errors in agreement with the Galster 
parameterization.
}


\section{Introduction}

In todays view the nucleon is composed of pointlike, almost massless
current quarks and gluons as the mediators of the color forces
between them.
The dynamics is described by the quantum field theory of strong 
interaction, QCD.
It is well tested at very high energies and momentum transfers,
where the mutual interaction is weak enough for perturbative
methods to be applicable.
The nucleon appears dramatically different at low energies 
(where we find our surrounding matter).
Here the non-linearity of QCD prohibits any exact {\it ab initio}
calculation of hadron properties.
This is the regime of constituent quarks as effective,
multi-body degrees
of freedom and the light mesons as the Goldstone bosons of
chiral symmetry breaking.
As for any composite, non--pointlike quantum mechanic particle 
\cite{Ericson73} a whole excitation spectrum
builds upon the nucleon ground state.
Although there remain decisive problems concerning the total number and 
individual nature of states,
the basic properties of the spectrum can be successfully reproduced by
constituent quark models \cite{CR00}.
However, the transition between the current and constituent quark regimes 
is not understood.
Lattice QCD and chiral extrapolations yield hints for the masses of
effective constituents \cite{HLT00},
but dynamically generated constituent quarks could not yet been identified
\cite{Wiese03}.
Nevertheless, first promising results have been obtained for ground state 
properties like magnetic moments, polarizabilities and form factors
\cite{Dong97}.

The intimate connection of ground state observables to the 
high energy structure is made explicit
by the concept of generalized parton distributions $H,\tilde H, E, \tilde E$
in exclusive deep inelastic reactions \cite{GPV01}.
At vanishing (Mandelstam) $t \rightarrow 0$ they fade to 
the usual unpolarized and polarized parton distributions
for quarks and antiquarks \cite{Radyushkin96},
\begin{equation}
H \rightarrow q(x) \qquad
\tilde H \rightarrow \Delta q(x),
\end{equation}
whereas in the non--perturbative regime of small but finite $t$
integration over Bj{\o}rken $x$ yields
the Dirac, Pauli and axial form factors \cite{Ji97},
\begin{eqnarray}
\int_{-1}^1 dx\,H &=& F_1(t) \qquad \int_{-1}^1 dx\,E = F_2(t) \\
\int_{-1}^1 dx\,\tilde H &=& g_A(t) \qquad \int_{-1}^1 dx\,\tilde E = h_A(t).
\end{eqnarray}

The elastic form factors parameterize the ability of the composite 
nucleon to incorporate a momentum transfer, $\vec q$, coherently, i.e.
without excitation and particle emission.
They are related to the distribution of charge and currents and therefore 
of fundamental importance for the understanding of nucleon structure.

Lepton scattering is a unique tool for the investigation of the 
electromagnetic structure of the nucleon.
In the simplest approximation, it is characterized  
by the exchange of one virtual photon, which transfers the momentum 
$\vec q$  and the energy $\omega$.
Electron scattering covers the spacelike region, because the squared
four-momentum transfer, $q^2 = \omega^2 - \vec q^{\,\,2}$, is always negative.
It is therefore usually expressed by the positive quantity 
$Q^2 = -q^2 > 0$.
The timelike region, where $Q^2 < 0$, can be accessed through 
electron-positron annihilation into a pair of 
proton and anti-proton \cite{Armstrong93} or neutron and anti-neutron
\cite{Antonelli98}. 


In elastic electron-nucleon scattering
the charge-current density of the nucleon can be written in the 
form \cite{BD64}:
\begin{equation}
\bar{N} \Gamma_\mu N = \bar{N} \left[ \gamma_\mu F_1(Q^2) +
              \frac{i \sigma_{\mu\nu} q^\nu}{2 m_N} \kappa F_2(Q^2) \right] N.
\end{equation}
The {\it Dirac} form factor, $F_1(Q^2)$,  modifies the vector current of 
charge and normal magnetic moment,
whereas the {\it Pauli} form factor, $F_2(Q^2)$, parameterizes the 
effect of the anomalous magnetic moment, $\kappa$, as motivated by the
Gordon decomposition of the electromagnetic current.
Linear combinations of $F_1$  and $F_2$  constitute the electric and 
magnetic {\it Sachs} form factors \cite{Sachs62}
\begin{eqnarray}
G_E^{n,p}(Q^2) &=& F_1^{n,p}(Q^2) - \tau\kappa_{n,p} F_2^{n,p}(Q^2) \\
G_M^{n,p}(Q^2) &=& F_1^{n,p}(Q^2) + \kappa_{n,p} F_2^{n,p}(Q^2),
\end{eqnarray}
where $\tau=Q^2/4m_N^2$  is a dimensionless measure of the squared 
four-momentum transfer in units of the nucleon rest mass, $m_N$.
At $Q^2 \rightarrow 0$ these form factors correspond 
to the total charge and magnetic moment of protons and neutrons:
\begin{eqnarray}
G_E^{n,p}(Q^2 \rightarrow 0) &=& 
     0, \,\,1 \\ 
G_M^{n,p}(Q^2 \rightarrow 0) &=& 
     -1.91,\,2.79\,.
\end{eqnarray}
In the particular reference frame with vanishing energy transfer, the 
{\it Breit} frame, $G_E$  and $G_M$  have been interpreted as the 
Fourier transforms of the corresponding distributions of charge and
magnetism \cite{Sachs62}.
Recently, the possibility of interpretation of $G_E$  in terms of the 
intrinsic charge distribution was controversially discussed 
again \cite{Isgur98,Cardarelli99,Leinweber01,Bawin01}.

With the Sachs form factors the cross section for elastic electron-nucleon
scattering can be written in the famous {\it Rosenbluth} form,
\begin{equation} 
\frac{d\sigma}{d\Omega} = \left( \frac{d\sigma}{d\Omega} 
                          \right)_{\mbox{{\small Mott}}} \cdot
                          \left(
                          \frac{G_E^2+\tau G_M^2}{1+\tau} +
                          2 \tau G_M^2 \tan^2\frac{\vartheta_e}{2}
                          \right),
\label{eq:Rosenbluth}
\end{equation}
where $\left(d\sigma/d\Omega \right)_{\mbox{\small Mott}}$  
is the Mott cross section for electron scattering 
off a pointlike spin--$\frac{1}{2}$  object and
$\vartheta_e$  denotes the electron scattering angle.
Due to their different angular weights in Eq.\ref{eq:Rosenbluth},
$G_E$  and $G_M$  can be experimentally separated at constant $Q^2$.

The measurement of all Sachs form factors of proton and neutron enables
the decomposition into isoscalar and isovector parts, 
important for the comparison to model calculations.
Furthermore, using isospin invariance it is possible to extract $u$/$d$
flavour specific form factors from $G_{E,M}^p$  and $G_{E,M}^n$. 
The extension to the electric and magnetic strange quark 
contributions requires additional observables.
From parity violating elastic electron scattering the cross section
asymmetry with regard to the helicity $+/-$ flip of the electron beam
\begin{equation}
A_{PV} = \frac{\sigma^+ - \sigma^-}{\sigma^+ + \sigma^-}
\end{equation}
is extracted, which is due to the interference of $\gamma$ and
$Z^0$  exchange.
It determines the $Z$ form factors
\begin{equation}
G_{E,M}^Z = G_{E,M}^p - G_{E,M}^n - 4 \sin^2\Theta_W\,G_{E,M}^p - G_{E,M}^s,
\end{equation}
from which the strange form factors $G_{E,M}^s$  can be extracted,
provided the Weinberg mixing angle, $\Theta_W$, and the nucleon form
factors $G_{E,M}^p$ and $G_{E,M}^n$ are sufficiently well known.

This talk reviews the current experimental status of the electromagnetic
elastic nucleon form factors. 
Recent results concerning the proton electric form factor at high $Q^2$
are summarized in the next section. 
The neutron form factors are discussed in section 
\ref{sec:neutron}. 
Meaurements of $G_M^n$ are addressed in subsection 
\ref{sec:neutron_magnetic}.
Special emphasis is given to recent measurements of the neutron electric
form factor using double polarization techniques in section 
\ref{sec:neutron_electric}.
The paper concludes with a summary and outlook.

\section{Proton form factors}
\label{sec:proton}

The method of Rosenbluth-separation has been used in elastic 
electron-proton scattering  \cite{Bosted92,Sill93} 
to determine $G_E^p$  and $G_M^p$  up to $Q^2 \simeq 9$\,(GeV/c)$^2$.
At higher $Q^2$  up to 30\,(GeV/c)$^2$, $G_M^p$  dominates the cross section
and has thus been determined directly \cite{Arnold86}.
The result was that both $G_M^p$ and $G_E^p$  approximately follow 
the so-called dipole form and scale with the magnetic moment, $\mu_p$:
\begin{eqnarray}
G_E^p &\simeq& G_D = \left( 1+\frac{Q^2}{0.71\,(\mbox{GeV/c})^2} \right) ^{-2}
\,,
\\
G_M^p &\simeq& \mu_p G_D.     
\end{eqnarray}

However, due to the insensitivity of the cross section to $G_E$ at higher
$Q^2$, the Rosenbluth separation yields large uncertainties
for this quantity.
Much improved sensitivity is obtained using double polarization
observables in 
electron-nucleon scattering \cite{AR74,ACG81}.
The components of the recoil polarization in the $N(\vec e, e'\vec N)$
reaction read
\begin{eqnarray}
P_x &=& - P_e \frac{\sqrt{2\tau\epsilon(1-\epsilon)} G_E^N G_M^N}
                   {\epsilon (G_E^N)^2 + \tau (G_M^N)^2}  
\label{eq:recpol_x}          \\
P_y &=& 0                                                       \\
P_z &=& P_e \frac {\tau\sqrt{1-\epsilon^2} (G_M^N)^2}
                  {\epsilon (G_E^N)^2 + \tau (G_M^N)^2},                
\label{eq:recpol_z}
\end{eqnarray}
where $\hat x$  is in the electron scattering plane 
perpendicular to the direction of the momentum transfer, 
$\hat y$  is normal to the scattering plane, and 
$\hat z$  points into the direction of the momentum transfer, $\vec q$;
$\epsilon = ( 1 + \frac{2|\vec q|^2}{Q^2} \tan^2{\frac{\vartheta_e}{2}} )^{-1}$
is the photon polarization parameter and
$P_e$  denotes the longitudinal polarization of the electron beam.
\begin{figure}[t]
\centerline{\epsfxsize=10cm\epsfbox{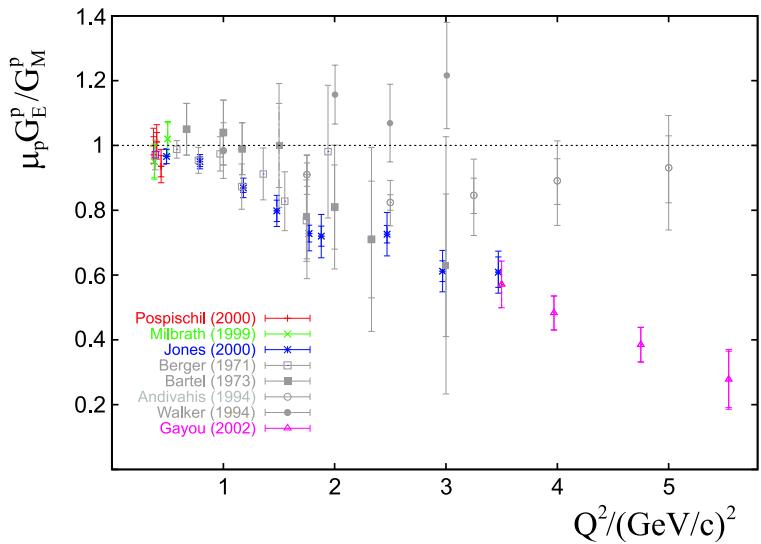}}   
\caption{Results for $\mu_p G_E^p/G_M^p$ from unpolarized Rosenbluth 
         \protect\cite{Berger71,Bartel73,Andivahis94,Walker94}
         and recoil polarization measurements 
         \protect\cite{Jones00,Gayou01,Gayou02,Milbrath99,Pospischil01}.}
\label{fig:GEp_results}
\end{figure}
In particular, $P_x$ is linear in $G_E$ and the polarization ratio
$P_x/P_z$ is directly related to $G_E/G_M$.

The measurements of the Hall A collaboration at Jefferson 
Laboratory using a recoil polarimeter in one of the high resolution 
spectrometers \cite{Jones97} show with high statistical precision
a linear decrease of $G_E^p/G_M^p$ up to $Q^2 = 5.6$\,(GeV/c)$^2$
\cite{Jones00,Gayou01,Gayou02} (c.f. Fig.\,\ref{fig:GEp_results}).
The results have been included in new empirical form factor fits
\cite{Brash02}. 
At low $Q^2$ recoil polarization measurements 
\cite{Milbrath99,Pospischil01} are in agreement
with form factor scaling, i.e. $\mu_p G_E^p/G_M^p = 1$.


\section{Neutron form factors}
\label{sec:neutron}

Generally, the measurement of the neutron form factors raises more 
difficulties, because there is no free neutron target available which is
suited for electron scattering experiments.
The best possible approximation of free electron-neutron scattering is
quasi--free scattering off the lightest nuclei.
In $D(e,e')$  
single arm experiments the dominating proton contribution has to be
subtracted, with corresponding large uncertainties \cite{Sick89}.
A Rosenbluth-separation has nevertheless been achieved \cite{Lung93} up to 
$Q^2=4$\,(GeV/c)$^2$.
At low $Q^2$  nuclear effects and final state interaction are well enough
under control \cite{Golak01} to permit an extraction of $G_M^n$  
from inclusive quasi--elastic $^3\vec{\mbox He}(\vec e,e')$
scattering with polarized beam and polarized target \cite{Gao94,Gao99,Xu00}.
$D(e,e'n)$  coincidence experiments allow the explicit tagging of 
electron-neutron scattering
\cite{Bartel73,Markowitz93,Anklin94,Bruins95,Anklin98,Kubon01}.
The influence of binding effects can be minimized through the simultaneous
measurement of the $D(e,e'p)$  reaction.
Similar to $G_M^p$, the neutron magnetic form factor also roughly exhibits 
the dipole behaviour, 
\begin{equation}
G_M^n \simeq \mu_n G_D.
\label{eq:dipole}
\end{equation}

The situation concerning the neutron electric form factor, $G_E^n$,  
is most unfavourable.
It must vanish in the static limit, $G_E^n (Q^2 \rightarrow 0) = 0$,
due to the zero charge of the neutron. 
The smallness of $(G_E^n)^2$  compared to $\tau (G_M^n)^2$  
makes a Rosenbluth decomposition according to
Eq.\ref{eq:Rosenbluth} very difficult.
Due to the corresponding large errors in the small quantity the extracted
values for $G_E^n$  are compatible with zero \cite{Bartel73,Lung93}.
Therefore, in the momentum transfer range $Q^2 < 1$\,(GeV/c)$^2$  the most
precise data came from {\it elastic} electron-deuteron 
scattering \cite{Galster71,Platchkov90},
where the structure function $A(Q^2)$  depends on the isoscalar 
form factor $(G_E^p + G_E^n)^2$  and thus provides a higher sensitivity
to $G_E^n$  through its interference with the large $G_E^p$.
However, the necessary unfolding of the deuteron wavefunction introduces
a substantial model dependence in the extracted neutron electric
form factor.
Reduced model dependence has been obtained analyzing the $e-d$ elastic
quadrupole form factor \cite{SS01}.
The model dependence can be overcome
by exclusive $(e,e'n)$  double polarization experiments
with polarized beam and either polarized target or recoil polarimetry.

\subsection{Recent $G_M^n$  experiments}
\label{sec:neutron_magnetic}

Precise determinations of $G_M^n$  come from unpolarized 
coincidence experiments at Bates \cite{Markowitz93}, 
NIKHEF \cite{Anklin94}, ELSA \cite{Bruins95}, and MAMI 
\cite{Anklin98,Kubon01}.
Except the Bates experiment, which measured the absolute $D(e,e'n)$  
cross section,
these experiments determined $G_M^n$  from the ratio 
$ R = \frac{\sigma(e,e'n)}{\sigma(e,e'p)} $
of quasi--free neutron over proton cross sections off the deuteron
with simultaneous neutron/proton detection in one single hadron detector.
In this ratio  
nuclear binding effects cancel to a large extent.
Moreover, this method is  also experimentally insensitive to 
luminosity fluctuations and detector acceptancies.

Nevertheless, detailed control of the hadron detectors
absolute detection efficiency for
protons and, particularly, neutrons is essential. 
The setups of the various experiments have been very similar, 
with the electrons
detected in a magnetic spectrometer and coincident n/p-detection in a well
shielded plastic or mineral oil scintillator telescope.
However, different approaches have been used to determine the absolute
neutron detection efficiency.
For the Bates and ELSA measurements "in situ" methods were chosen 
with a bremsstrahlung radiator postioned in front of the experimental target
in order to exploit the $D(\gamma,p)n$  or
$p(\gamma,\pi^+)n$  reactions, respectively, to tag neutrons in the telescope.
In contrast, the hadron detectors which were used at MAMI and NIKHEF 
were calibrated at the PSI neutron beam in a kinematically complete
$p(n,p)n$  experiment.
This method relies on the good control and portability of the effective
detector thresholds.

\begin{figure}
\centerline{\epsfxsize=10cm\epsfbox{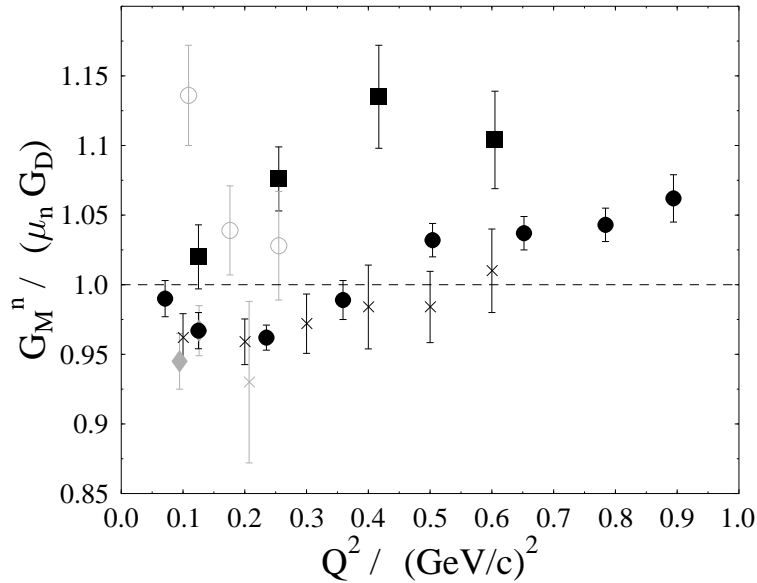}}   
\caption{Results of recent $G_M^n$  experiments. 
         Gao et al. {\protect\cite{Gao94,Gao99}} (grey cross) and
         Xu et al. {\protect\cite{Xu00,Xu02}} (black crosses) exploited the 
         $^3\vec{He}(\vec e, e')$ cross section asymmetry. 
         The open circles are the result of the D$(e,e'n)$  experiment
         of Markowitz et al.{\protect\cite{Markowitz93}}. 
         The ELSA {\protect\cite{Bruins95}} (full squares),
         NIKHEF {\protect\cite{Anklin94}} (grey diamonds) and
         MAMI {\protect\cite{Anklin98,Kubon01}} (full circles) measurements
         made use of the ratio method as described in the text.}
\label{fig:gmn}
\end{figure}

Figure\,\ref{fig:gmn} shows a comparison of the recent results from 
Bates \cite{Gao94,Gao99,Markowitz93}, ELSA \cite{Bruins95},
JLab \cite{Xu00,Xu02},
NIKHEF \cite{Anklin94}, and MAMI \cite{Anklin98,Kubon01}.
Despite the individual errors of down to 2\,\%, the MAMI data
are approximatly 10\,-\,15\,\% below the ELSA ones. 
The probable origin of this discrepancy is the absolute neutron detection
efficiency calibration.
In this respect, the possible impact of un-tagged electroproduction events 
has been discussed for the in situ method \cite{Jourdan97,Bruins97}.

The measurement of the transverse asymmetry $A_{T'}$  from inclusive
quasi--elasic $^3\vec{\mbox He}(\vec e,e')$  scattering provides an alternative
method for the extraction of $G_M^n$, which is completely independent of
efficiency calibrations \cite{Gao94,Gao99,Xu00}.
However, full Fadeev calculations are required with inclusion of final state 
interaction and meson exchange currents.
Those are currently available for $Q^2 = 0.1$  and $0.2$\,(GeV/c)$^2$
\cite{Golak01,Xu00}, 
the results of a PWIA analysis for $Q^2$ up to 
$0.6$\,(GeV/c)$^2$ \cite{Xu02}.
The extracted values of $G_M^n$  agree with the unpolarized measurements of
Anklin et al. \cite{Anklin94,Anklin98} and Kubon et al.\cite{Kubon01}.

\subsection{$G_E^n$ double polarization experiments}
\label{sec:neutron_electric}

Double polarization observables in exclusive quasi--free electron-deuteron 
scattering with longitudinally polarized electrons offer high sensitivity to 
$G_E^n$, due to an interference with the large $G_M^n$, 
combined with negligible dependence on the deuteron wavefunction
\cite{Arenhoevel87-88},
e.g. in the recoil polarization observables of Eqs.\,\ref{eq:recpol_x}
and \ref{eq:recpol_z}. 
%
In the completely equivalent scattering $\vec n(\vec e,e'n)$ of longitudinally 
polarized electrons off a polarized neutron target
the cross section asymmetry with regard to reversal of the electron beam
polarization is given by
\begin{equation}
A = -P_e \frac{ \sqrt{2\tau\epsilon(1-\epsilon)} G_E^n G_M^n \cdot \tilde{P_x}
                + \tau\sqrt{1-\epsilon^2} (G_M^n)^2 \cdot \tilde{P_z} }
              { \epsilon (G_E^n)^2 + \tau (G_M^n)^2 },
\label{eq:asymmtry}
\end{equation}
where now $\tilde{P}_{x,z}$  are the components of the initial state
neutron polarization.
The polarized target neutrons can be provided by polarized 
$\vec{^3\mbox{He}}$  \cite{Eckert92,Woodward90,Chupp92},
where the neutron carries approximately 87\,\% of the 
polarization of the nucleus \cite{Friar90},
or by vector polarized $^2 \vec D$  \cite{Passchier99,Zhu01}.
The measurement of asymmetry ratios yields independence of
the absolute degree of the target polarization 
\cite{Meyerhoff94,Passchier99}.

\subsubsection{Polarized target experiments}

First experiments aimed at the extraction of $G_E^n$
from the inclusive quasi--elastic reaction $\vec{^3\mbox{He}}(\vec e, e')$.
The feasibility of such kind of experiments was
successfully demonstrated, 
but the statistical accuracy remained unsatisfactory.

The magnetic moment of $^3\mbox{He}$  within 10\,\% agrees with the free 
neutron one. 
Therefore the proton contribution in the measured asymmetries 
first was expected to be small \cite{Friar90}.
Later calculations, however, showed that the remaining impact of
the protons on the measured asymmetries is large enough to prohibit
a reliable extraction of $G_E^n$  \cite{Schulze-Sauer93}.

This problem can be overcome, if the occurence of e-n scattering is
explicitly tagged through the detection of the outgoing neutron in
coincidence with the scattered electron. 
Such an exclusive $^3\vec {\mbox{He}} (\vec e, e'n)$  
experiment was performed for the 
first time by the A3-collaboration at MAMI \cite{Meyerhoff94} at a
squared four-momentum transfer of $Q^2 = 0.31$\,(GeV/c)$^2$. 
With the detector-setup described in the following section the 
statistics was improved later on \cite{Becker99}.
The most recent experiment at $Q^2 = 0.67$\,(GeV/c)$^2$
used the 3-spectrometer setup \cite{Blomqvist98}
of the A1-collaboration at MAMI \cite{Rohe99,Bermuth01}.

In this experiment the target gas was polarized by metastable optical pumping 
and subsequently compressed to 6 bars.
The relaxation time of approximately one day required twice a day
the replacement of the target cell by a freshly polarized one.
Quasi--elastic measuremets were performed with target spin aligned
perpendicular and parallel to the momentum transfer direction in order to
access both the transverse asymmetry, $A_x$, and the longitudinal asymmetry,
$A_z$.
This enabled a measurement of the ratio $A_x/A_z$, which is directly
proportional to $G_E^n / G_M^n$  but independent of both the absolute
degrees of beam and target polarization, $P_e$  and $P_T$, respectively.
Furthermore, the product $P_e \cdot P_T$  was monitored through the
{\it elastic} measurement $^3\mbox{He} (e, e')$  in spectrometer B
of the 3-spectrometer setup.

\begin{figure}
\centerline{\epsfxsize=10cm\epsfbox{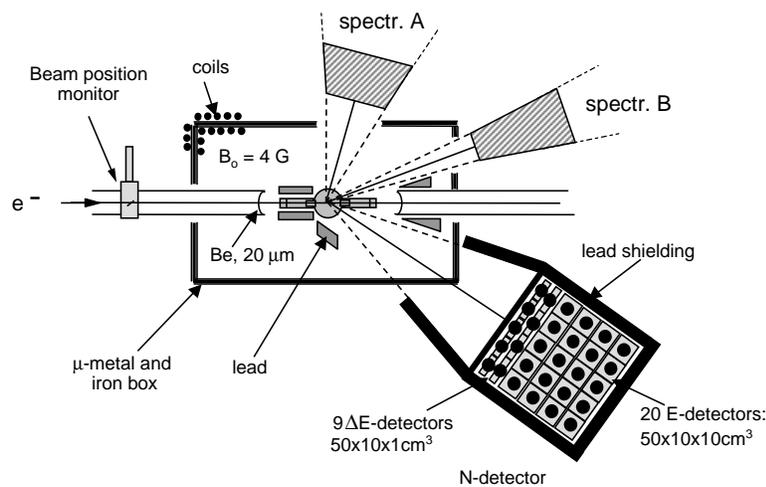}}   
\caption{Target area of the $^3\vec {\mbox{He}} (\vec e, e'n)$  
         experiment at MAMI. The cell with polarized gas is magnetically
         shielded against the spectrometers fringe fields.}
\label{fig:3-He_setup}
\end{figure}


The quasi--elastically scattered 
electrons were detected  in spectrometer A (c.f. Fig.\ref{fig:3-He_setup}), 
the neutrons in coincidence in
a dedicated neutron detector provided by the University of Basel.
It consisted of four layers of five plastic scintillators of dimensions
$50 \times 10 \times 10$\,cm$^3$, which were equipped with photomultipliers at
both ends. 
Charged particles could be rejected by means of
two layers of 1\,cm thick $\Delta$E-counters. 
The neutron detector was shielded with 2\,cm of lead against direct 
target sight in order to reduce the charged background.

The setup of the JLab Hall C $\vec D (\vec e,e'n)$  experiment in principle is 
similar \cite{Day93,Zhu01}.
Here the scattered electrons are detected in the HMS spectrometer in 
coincidence with neutrons in a segmented plastic scintillator.
At the required luminosity of $1 \cdot 10^{35}$\,cm$^{-2}$s$^{-1}$
a 40\% polarization of the ND$_3$  target is achieved 
by the technique of dynamic nuclear polarization. 
The deuteron nuclei are polarized by microwave irradiation 
at temperatures around 1\,K in a strong magnetic field of 5\,T.
In the measurement of $A_x$  this field deflects the incoming electron
beam by as much as $4^{\circ}$.
This has to be compensated by a magnetic chicane in order to guarantee
horizontal beam at the center of the target.
Data have been taken in the $Q^2$  range between $0.5$  and
2\,(GeV/c)$^2$, first results are available at the lowest $Q^2$  
\cite{Zhu01}.

In contrast to MAMI and JLab, the NIKHEF $\vec D (\vec e,e'n)$  experiment
was performed with a vector polarized {\it internal} gas target at the AmPS
electron storage ring \cite{Passchier99}.
The scattered electrons were detected in coincidence simultaneously with
protons and neutrons. 
Thus the asymmetry ratio between the $\vec D (\vec e,e'n)$  and 
$\vec D (\vec e,e'p)$  reactions could be determined.

\subsubsection{The \Den recoil polarization experiment at MAMI}

As in the case of polarized targets a pioneering recoil polarization
experiment was performed at MIT-Bates \cite{Eden94}.
Electrons and neutrons from the $D(\vec e, e'\vec n)$  reaction were detected 
in coincidence
and the transverse neutron polarization, $P_x$, was measured.
However, due to the low duty cycle of the Bates linac only modest statistical
accuracy could be achieved.
Furthermore, the external absolute calibration of the neutron polarimeter's
effective analyzing power remained unsatisfactory. 

\begin{figure}[ht]
\centerline{\epsfxsize=10cm\epsfbox{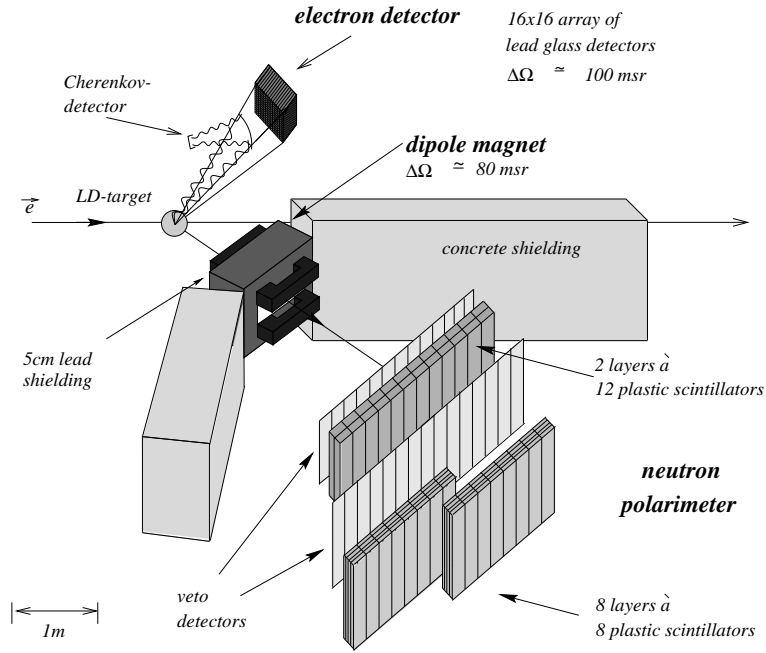}}   
\caption{Setup of the $D(\vec e, e' \vec n)$  experiment at MAMI}
\label{fig:setup}
\end{figure}

\begin{figure}[ht]
\centerline{\epsfxsize=10cm\epsfbox{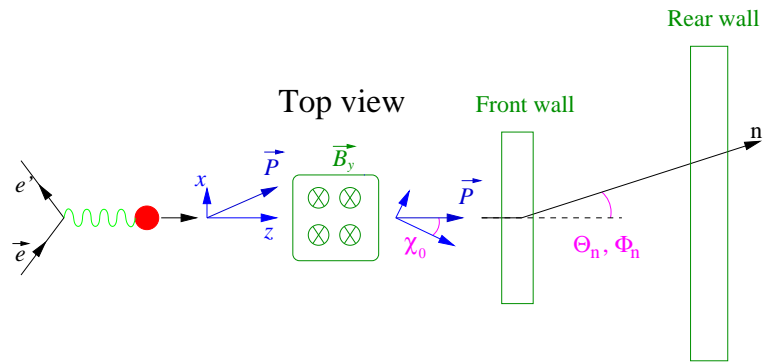}}   
\caption{Schematics of the spin precession}
\label{fig:precession_method}
\end{figure}

The full potential of the recoil polarization method was exploited for 
the first time at MAMI.
Fig.\ref{fig:setup} shows the experimental setup of this experiment.
The longitudinally polarized electron beam 
($I \simeq 2.5\,\mu$A, $P_e \simeq 75$\,\%)
impinged on a 5\,cm long liquid deuterium target
and the scattered electrons were detected in a 256 element lead glass 
array, 
which covered a solid angle of $\Delta\Omega \simeq 100$\,msr. 
The energy resolution of $\delta E / E \simeq 25$\,\% was sufficient to
suppress pion production events. 
Only the electron angles, which were measured with an accuracy of 
approximately $3.5$\,mrad entered the 
event reconstruction,
which became kinematically complete through the measurement of the
neutrons time-of-flight and hit position in the front plane of
the neutron detector.
\begin{figure}[ht]
\centerline{\epsfxsize=10cm\epsfbox{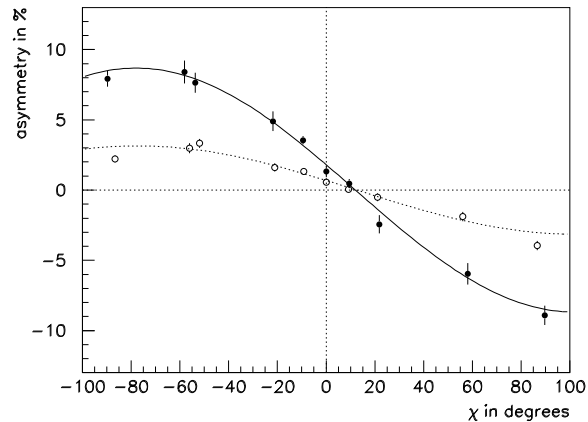}}   
\caption{Transverse asymmetries as a function of the spin precession angle. 
         The zero crossing is not affected by the different effective 
         analyzing powers obtained by different conditions in the offline
         analysis (full and open points).} 
\label{fig:asym}
\end{figure}

The neutron polarization can be analyzed in the detection process 
itself \cite{Taddeucci85}.
This required a second neutron detection in one of 
the rear detector planes,
which yielded the polar and azimuthal angles, $\Theta_n'$  and $\Phi_n'$,
of the analyzing scattering in the front wall.
With the number of events $N^{\pm}(\Phi_n')$  for $\pm$  helicity states of
the electron beam the azimuthal asymmetry, $A(\Phi_n')$, 
was determined through the ratio
\begin{equation}
\frac{1-A(\Phi_n')}{1+A(\Phi_n')} =
   \sqrt{ \frac{ N^+(\Phi_n') \cdot N^-(\Phi_n'+\pi) }
               { N^-(\Phi_n') \cdot N^+(\Phi_n'+\pi)} } ,
\end {equation}
which is insensitive to variations of detector efficiency and luminosity.
The extraction of $P_x$  from 
$A(\Phi_n') = \epsilon_{\mbox{eff}} \cdot P_x \cdot \sin\Phi_n'$  
requires the calibration of the effective analyzing power, 
$\epsilon_{\mbox{eff}}$, of the polarimeter.
This, however, varies strongly with the event composition as determined
by hardware conditions during data taking and software cuts applied in
the offline analysis.

The problem of calibration of the effective analyzing 
power has been avoided by controlled precession of the neutron spins in the
field of a dipole magnet in front of the polarimeter
\cite{Ostrick99}.
This is schematically depicted in Fig.\,\ref{fig:precession_method}.
After precession by the angle $\chi$  the transverse neutron polarization
behind the magnet, $P_{\perp}$, is a superposition of $x$  and $z$
components, and likewise is the measured asymmetry:
\begin{equation}
A_{\perp} = A_x \cos\chi - A_z\sin\chi.
\end{equation}
In the particular case of the zero crossing, $A_{\perp}(\chi_0) = 0$,
one immediately gets the relation
\begin{equation}
\tan\chi_0 = \frac{A_x}{A_z} = \frac
{\epsilon_{\mbox{eff}} \cdot 
 P_e\cdot\sqrt{2\tau\epsilon(1-\epsilon)}\,G_E^n\cdot G_M^n}
{\epsilon_{\mbox{eff}} \cdot 
 P_e\cdot\tau\sqrt{1-\epsilon^2}\,(G_M^n)^2 }.
\end{equation}
Obviously, this ratio is independent of both the degree of electron 
beam polarization, $P_e$, and the polarimeter's effective analyzing power.
It therefore directly yields $G_E^n/G_M^n$.
Different effective analyzing powers do change the magnitude of the transverse
asymmetry, $A_{\perp}$,  but not the zero crossing angle, 
$\chi_0$  \cite{Ostrick99}.
This is demonstrated in Fig.\,\ref{fig:asym}, where $A_{\perp}$  is plotted
for two different cut conditions in the offline analysis (full and open points)
as a function of the spin precession angle, $\chi$.
The magnitude of the asymmetry is affected but the zero crossing remains
unchanged. 

Using the established method of neutron spin precession a very similar 
experiment of the A1 collaboration at MAMI covers the extended $Q^2$  range
up to $0.8$\,(GeV/c)$^2$ \cite{HS99,Seimetz_dr,Glazier_dr}.
Given the maximum beam energy of 880 MeV this requires neutron detection 
at forward angles of $\Theta_n^\text{lab} \simeq 27^\circ$ in a highly 
segmented neutron polarimeter.

\subsection{Results}
\label{sec:results}

Even for quasi--free $D(e, e'n)$ scattering FSI effects are substantial below
$Q^2 \simeq 0.3$\,(GeV)$^2$  and have been taken into account 
in the recent analyses
using the calculations of Arenh{\"o}vel \cite{Arenhoevel87-88}.
They drop rapidly with increasing $Q^2$, increasing $G_E^n$ for the
MAMI/A3 data sample by almost a factor of two at $Q^2 = 0.12$\,(GeV)$^2$ 
but only $\simeq 10\,\%$  at $0.32$\,(GeV)$^2$  \cite{Herberg99}.
Despite its size, the required correction of the $G_E^n/G_M^n$  ratio
has only
small uncertainties, because it is insensitive to the choice
of N-N potential and $G_E^n$  parameterization 
\cite{Arenhoevel87-88,Arenhoevel98}.
Therefore, even at $Q^2 = 0.12$\,(GeV/c)$^2$  a reliable extraction of 
$G_E^n/G_M^n$  is possible.
This has been done relying on the dipole values for $G_M^n$  
(Eq.\ref{eq:dipole}).

\begin{figure}[t]
\centerline{\epsfxsize=8cm\epsfbox{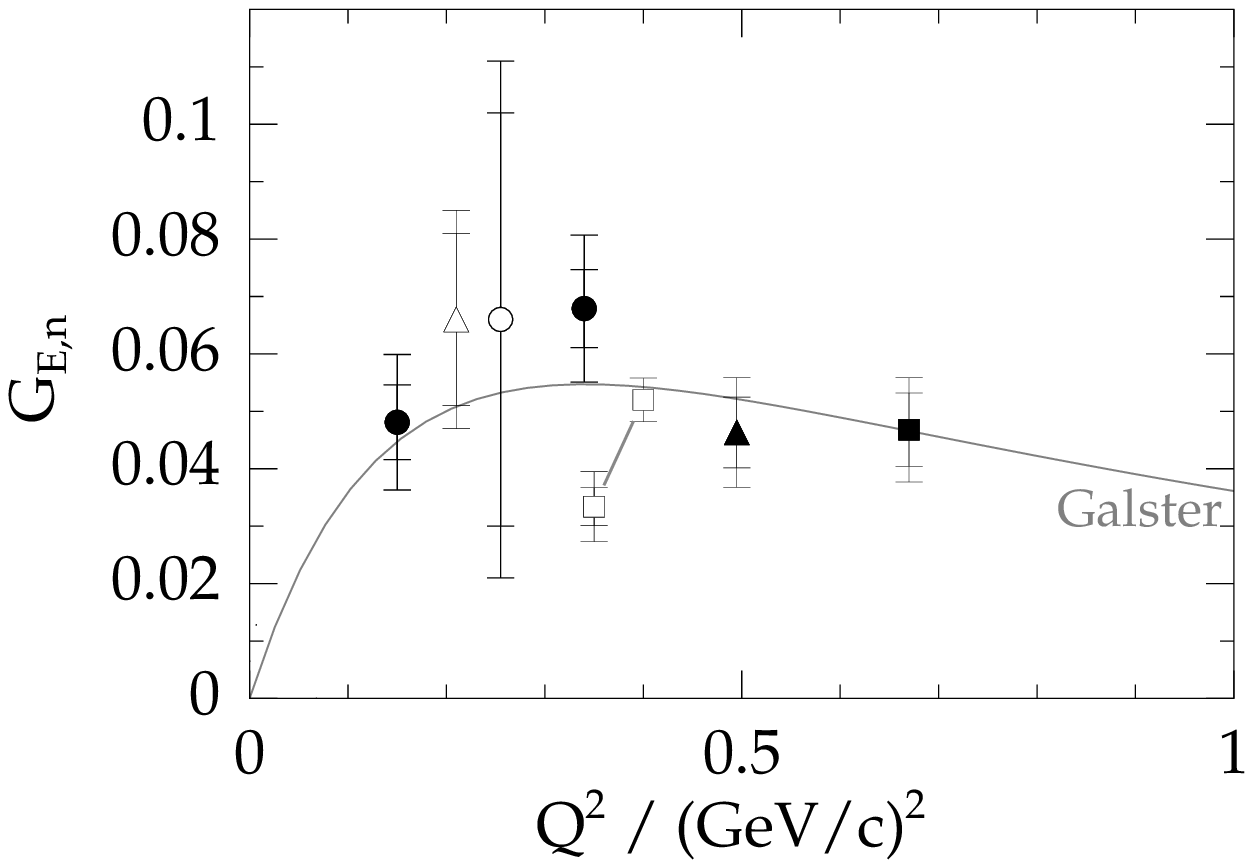}}   
\caption{Results for $G_E^n$  from the \Den experiment at MAMI
         (full circles) {\protect\cite{Ostrick99,Herberg99}} 
         along with the \Hen results from MAMI 
         (full square {\protect\cite{Rohe99,Bermuth01}}, open squares 
         \protect\cite{Becker99} with and without FSI correction), 
         the \Den result from Bates 
         (open circle) {\protect\cite{Eden94}},
         and the $\vec D(\vec e,e'n)$  results from NIKHEF
         (open triangle {\protect\cite{Passchier99}}) 
          and
         JLab (full triangle {\protect\cite{Zhu01}}).
         Except the Bates data point the deuterium results are FSI-corrected.
         The curve is the one--parameter Galster parameterization.
         \protect\cite{Galster71}}
\label{fig:gen}
\end{figure}
Fig.\,\ref{fig:gen} gives a summary of the recent double polarization 
measurements with statistical (inner) and systematical (outer) errors.
The recoil polarization experiments are depicted by circles,
full for the MAMI results \cite{Ostrick99,Herberg99} and
open for the Bates one \cite{Eden94}.
The triangles indicate the NIKHEF \cite{Passchier99} and
JLab \cite{Zhu01} measurements with polarized deuteron target.
The squares represent the MAMI results for \Hen 
\cite{Becker99,Rohe99,Bermuth01}. 
For the $^3\vec {\mbox He}$  experiments FSI is expected to be small
at $Q^2 = 0.67$\,(GeV/c)$^2$  \cite{Rohe99,Bermuth01}, 
due to the large kinetic energy of the ejected neutron.
Contrary, at $Q^2 = 0.36$\,(GeV/c)$^2$  (open squares) \cite{Becker99} 
first, still incomplete, Faddeev calculations 
indicate a substantial correction of the \Hen data point
towards larger $G_E^n$.
At the present status of calculation where no meson exchange currents are
yet included the central value of the extracted $G_E^n$  is shifted by 
approximately 50\,\% \cite{Golak01}.
Due to the kinematical reconstruction the average $Q^2$  is also affected.
Despite the small statistical error, this data point is subject to
the largest systematical and theoretical (model) uncertainty,
which is not included in the depicted error.

All polarization data lie above the so far favoured result from
elastic $D(e,e')$  scattering, where the Paris potential has been used for the
unfolding of the wave function contribution \cite{Platchkov90}.
They are compatible with the older Galster parameterization \cite{Galster71}
\begin{equation}
G_E^n = - \frac{\mu_n \tau}{1 + \eta \tau} \cdot G_D
\end{equation}
with $\eta = 5.6$, which is indicated by the line in Fig.\,\ref{fig:gen}.

\section{Summary and Outlook}
\label{sec:summary}

Elastic form factors are related to the distribution of charge
and current and thus fundamental quantities characterizing 
the nucleon ground state.
In the context of this conference they are a prerequisite for the
extraction of strangeness--specific information from parity violating
elastic electron scattering.

From single arm $e-p$ scattering
precise proton magnetic form factor data are available up to 
$Q^2 = 30$\,(GeV/c)$^2$, roughly exhibiting dipole behaviour.
Electric and magnetic contributions have been separated using the 
Rosenbluth technique;
the results are in agreement with with $p(\vec e, e'\vec p)$ measurements
at low $Q^2$.
Above $Q^2 \simeq 1.5$\,(GeV/c)$^2$ the double polarization experiments 
have shown the ratio $\mu_p G_E^p/G_M^p$ to linearly fall below unity,
$G_E^p$  remaining only about 30\,\% of the dipole value at 
$Q^2 = 5.6$\,(GeV/c)$^2$.

The neutron magnetic form factor, $G_M^n$, could be extracted up to 
$Q^2 = 4$\,\,(GeV/c)$^2$  from single arm $D(e,e')$  quasi--elastic scattering.
Below $Q^2 = 1$\,\,(GeV/c)$^2$, recent coincidence experiments allowed a
precise determination of $G_M^n$  from the ratio 
$R = \frac{\sigma(e,e'n)}{\sigma(e,e'p)}$  
of quasi--free electron scattering cross sections off the deuteron. 
Despite their individual statistical errors of only 2\,\%,
two datasets from ELSA and MAMI/NIKHEF differ by as much as 10 - 15\,\%.
This discrepancy contributed a substantial error to 
the extraction of strange form factors from parity violation. 
Recent independent results from $^3\vec {\mbox{He}} (\vec e, e'n)$  
measurements at $Q^2 \leq 0.6$\,(GeV/c)$^2$  
agree with the MAMI/NIKHEF dataset, 
however partially still relying on a simplified PWIA analysis.
The full solution of this problem is also important 
with regard to the normalization
of the recent double polarization experiments \Hen,
$\vec D(\vec e,e'n)$   and \Den  at various laboratories,
where the neutron electric form factor is extracted from the ratio 
$G_E^n/G_M^n$  of electric to magnetic neutron form factors.

In the \Den experiment at MAMI the neutron polarimeter was 
supplemented by a spin precessing dipole magnet. 
This technique, which is also adopted for the corresponding JLab experiment, 
avoids the polarimeters analyzing power calibration.
The influence of final state interaction has been quantitatively evaluated
for the \Den reaction.
All recent double polarization experiments are compatible with the
old Galster paramerization.

It will be important to continue these $G_E^n$  experiments at larger 
$Q^2$.
At JLab and MAMI there are further measurements underway, both with 
polarized target \cite{Day93,Heil95} 
and recoil polarimetry \cite{Madey93,HS99}.
These experiments will further exploit the potential of double polarized
quasi--elastic electron scattering.

\section{Acknowledgements}

The MAMI experiments have been performed within the framework of the
collaborations A1 and A3 with contributions of institutes of the 
Universities of Basel, Bonn, Glasgow, Mainz, and T\"ubingen.
H. Arenh\"ovel supplied all the necessary calculations 
for the FSI correction of the deuteron measurements.
The spin precession magnet for the \Den experiment was provided by 
the Physikalisches Institut of the University of Bonn.
The MAMI experiments are financially supported by the 
Deutsche Forschungsgemeinschaft (SFB\,443). 


\end{document}